# Secure Communication Using Electronic Identity Cards for Voice over IP Communication, Home Energy Management, and eMobility

Rainer Falk, Steffen Fries, Hans Joachim Hof
Corporate Technology
Siemens AG
Munich, Germany
e-mail: [rainer.falk | steffen.fries | hans-joachim.hof]@siemens.com

*Abstract*—Using communication services is a common part of everyday life in a personal or business context. Communication services include Internet services like voice services, chat service, and web 2.0 technologies (wikis, blogs, etc), but other usage areas like home energy management and eMobility are will be increasingly tackled. Such communication services typically authenticate participants. For this identities of some kind are used to identify the communication peer to the user of a service or to the service itself. Calling line identification used in the Session Initiation Protocol (SIP) used for Voice over IP (VoIP) is just one example. Authentication and identification of eCar users for accounting during charging of the eCar is another example. Also, further mechanisms rely on identities, e.g., whitelists defining allowed communication peers. Trusted identities prevent identity spoofing, hence are a basic building block for the protection of communication. However, providing trusted identities in a practical way is still a difficult problem and too often application specific identities are used, making identity handling a hassle. Nowadays, many countries introduced electronic identity cards, e.g., the German "Elektronischer Personalausweis" (ePA). As many German citizens will possess an ePA soon, it can be used as security token to provide trusted identities. Especially new usage areas (like eMobility) should from the start be based on the ubiquitous availability of trusted identities. This paper describes how identity cards can be integrated within three domains: home energy management, vehicle-2-grid communication, and SIP-based voice over IP telephony. In all three domains, identity cards are used to reliably identify users and authenticate participants. As an example for an electronic identity card, this paper focuses on the German ePA.

*Keywords - eMobility security; home energy management security; VoIP security; Smart Grid security; electronic identity card; elektronischer Personalausweis; authentication; identification*

## I. INTRODUCTION

Communication services use identifiers to indicate the intended recipient of a message or to setup a call. In the old telephone system, a telephone number according to ITU-T E.164 was used. But further identifiers are used nowadays for communication, e.g., email addresses, URLs of personal Web page, SIP URIs, Network Access Identifiers (NAI), customer numbers, or instant messaging screen names.

The SIP protocol is used as signaling protocol for Voice over IP (VoIP) communication and is also the base for different instant messaging applications. It uses a SIP URI to identify a user, both the originating and destination party.

The increasing usage of VoIP leads also to the fact that annoyance by unsolicited communication is not restricted to email SPAM, but also to voice calls. Such unsolicited communication is called SPIT, SPAM over Internet Telephony. Some ad hoc countermeasures include the filtering of incoming calls using a whitelist of permitted callers, a blacklist of denied callers, or to make a decision based on the caller's reputation that may be obtained from a reputation system. Also security-critical usages of voice communication take place commonly, as e.g., giving a bank order over a telephone line, or requesting information from a public administration office about the own case.

However, in all these cases trustworthy information about the identity of the communication partner is required as well as the verification of the authenticity of the identity information by either the communication peer or an intermediate component, vouching for the identity of the caller/calleé. Otherwise, identities can be spoofed e.g., to circumvent whitelists and blacklists. Another issue is that obtaining identities must involve some costs; otherwise a malicious party can simply change its identifier whenever it has bad reputation or is blacklisted.

The SIP standard defining VoIP signaling supports various authentication options that apply certain identifiers. Beyond them are direct authentication options but also assertions issued by a trusted third party. However, broad deployment of such a security solution requires a common security infrastructure for identity attestation. As the deployment of a new security infrastructure is costly, it is attractive to reuse an already existing security infrastructure for VoIP/SIP security. With the appearance of electronic identity cards as, e.g., the "Elektronischer Personalausweis (ePA)" in Germany that provides functionality for authentication towards an online service and trusted identities, such a security infrastructure is readily available. Moreover, the security architecture introduced by an official identity card is likely to be considered trustworthy by many people as the identity card is issued by the government and follows defined rules from the initial user authentication till the final identity card emission. Integration of ePA into VoIP communication has already been described in short in [1]. This paper presents the integration of identity cards in mature applications like VoIP communication as well as in upcoming applications like home energy management, and eMobility. The German "Elektronischer Personalausweis" ePA is taken as an example for how an identity card can be used for authentication of communication partners and how trusted identities of the identity card can be applied in the communication.

The remainder of this paper is structured as follows: Section II introduces ePA user authentication and ePA web authentication as examples of the authentication mechanisms of an identity card that is issued by a trusted instance (the German government). Section III gives an overview of authentication and identification in VoIP communication. Section IV describes identity handling in SIP. Section V describes different VoIP use cases that would profit from an integration of the authentication mechanisms of the "Elektronischer Personalausweis" into the SIP protocol. Section VI and VII describes the use of the "Elektronischer Personalausweis" in the usage area home energy management and eMobility respectively. Technical approaches and practically viable options for integration ePA-based user authentication within SIP/VoIP communication, home energy management, and eMobility are described. Section VIII provides an outlook, while Section IX concludes this paper.

## II. AUTHENTICATION USING THE ELECTRONIC IDENTITY CARD EPA

The authentication function of the ePA allows secure transmission of attributes from the ePA to a third party. Attributes may be related to a person (name) or to a property or characteristic (e.g., age). Even relative attributes like "holder is older than 18" are possible. Attributes related to a person may be used as identities in VoIP communication. The corresponding authentication mechanism hence allows identification of communicating parties. The ePA offers various forms of authentication. In the following, the Extended Access Control is described:

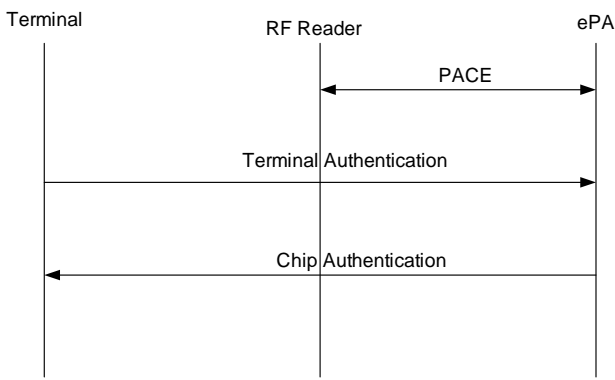

Figure 1. PACE Message Flow

Extended Access Control between a Terminal and an ePA involves three phases, see Figure 1: In the first phase, the PACE protocol is executed that is necessary to access the ePA. In the second phase, the terminal authenticates against the ePA, and in the third phase the ePA authenticates against the Terminal.

- PACE: PACE is a password authenticated Diffie-Hellman key exchange. A session key for protection of the communication is set up between the Radio Frequency (RF) Reader and the ePA.
- Terminal authentication: Each terminal has a terminal certificate for identification. The certificate is signed by a root CA. All root CA certificates are based on an international public key infrastructure (PKI). The German root CA is hosted by the BSI (Bundesamt für Sicherheit in der Informationstechnologie). Terminal certificates have a very short lifetime (24 hours). However, the ePA does neither have a certificate revocation list nor a physical clock. Instead, it stores the time of the last successful verification and does only accept timestamps that lie in the future. The terminal certificate includes information encoding which attributes of the user may be provided to that terminal.
- Chip authentication: A Diffie-Hellman authentication using a static chip key is performed to authenticate the ePA towards the Terminal.

Another authentication function of the ePA is web authentication (cf. Figure 2).

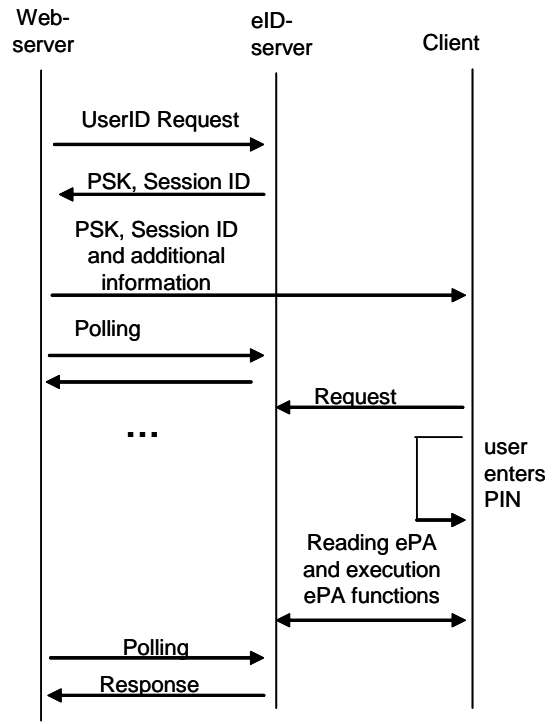

Figure 2. Web Authentication

A web server requests authentication of a user by sending a UserID Request to the eID server. The eID server provides a symmetric key (PSK) as well as a Session ID to the web server. The web server sends the PSK, the Session ID, and additional information related to the authentication request to the client of the user. After sending this message, the web server starts polling the eID server for a response of the authentication request. Upon reception of information from the web server, the client of the user sends an authentication request to the eID server. This starts the authentication between the eID server and the web client of the user. In the process of authentication, the user enters its PIN to activate the ePA. The eID server reads the ePA and executes ePA function. When this process finished, the eID server sends a response upon reception of a polling message of the web server. The response includes the result of the authentication request. The provider of the web server must ensure that the confidentiality of the PSK is protected during transfer, e.g. by using TLS/SSL, see [10] for details. Nowadays, eID services are offered by various providers, e.g. Bremen Online Services GmbH [11], Bundesdruckerei GmbH [12], and Deutsche Post Com GmbH [13]. Those providers offer integration modules for use in web servers.

This authentication approach can be integrated in existing multimedia applications as well as be used for upcoming usage areas as shown in the next sections. Keying material established by the authentication mechanism (some authentication methods support inband key agreement) may be used for further protection of, the established communication channel (e.g., for protection of integrity and confidentiality of VoIP communication, which is out of scope of this paper).

### III. AUTHENTICATION AND IDENTIFICATION IN VoIP COMMUNICATION

Main security objectives with relevance for VoIP communication concern the authentication and identification of participants, and the protection (integrity, confidentiality) of signaling and media data. Authentication and identification are the basis for the protection of VoIP communication; hence the focus of this paper is on authentication and identification. This section highlights some basic concepts of authentication and identification of participants in VoIP communication.

One may distinguish device authentication and user authentication: device authentication authenticates devices, even if they are shared between different users. Standard voice communication often uses device authentication, e.g., a telephone that is available to more than one person but only has one telephone number (used as identifier) associated. User authentication in contrast authenticates users, hence distinguishes between different users even if they use the same device. Appropriate identifiers for the intended authentication target need to be used. Device authentication and user authentication may be used together, or between different parties, e.g., device authentication to the service provider and user authentication to the calleé.

Another important point is who authenticates to whom. One may distinguish here between unilateral authentication and mutual authentication. For example, in unilateral authentication, a caller authenticates to a service provider. Mutual authentication is used if the service provider also authenticates to the caller. Mutual authentication may be necessary to avoid certain types of attacks. One example may be Man-in-the-Middle attacks, were the attacker manages to be an intermediate node in the communication path between the user and the server acting as the server towards the user and misusing the user credentials towards the server.

The endpoints of authentication are also of interest: the caller may authentication to a service provider or to the calleé. In the case of mobile phones, a user authenticates to the service provider and no end-to-end authentication between caller and calleé takes place. This approach is based on the trust users have in their service providers. The service provider may or may not assert the callers identity to the calleé. Asserted identities by a service provider may be trustworthy if only one service provider is involved (e.g., if a mobile phone connects to another mobile phone in his own network). However, if a call involves several service providers, the trustworthiness of the asserted identity may be much lower. Even worse, if the service provider use different technologies like one uses VoIP and the other one PSTN, it is more likely that such information is not provided end-to-end.

### IV. IDENTITY HANDLING IN SIP

The SIP protocol is the major signaling protocol for VoIP communication. It can be used for instant messaging (SIMPLE) as well. The SIP protocol defined in RFC3261 [2] supports two basic options for providing identity information in the SIP header to peers and also several options for authenticating users. For the provisioning of identity information the *From* header field defined as part of RFC3261 can be used as well as the *P-Asserted-Identity* header field, which is defined in RFC3325 [7]. While the first identity field is provided by the client itself, the latter one is typically used between trusted intermediaries (proxies). The following authentication options utilizing at least one of the fields named, but may not always provide true end-to-end protection of these fields. This is due to the fact that intermediate entities may alter some of the fields. This can be done for instance through Back-to-Back User Agents, which terminate a session setup in both directions. These entities are neither explicitly defined by SIP nor forbidden.

#### A. SIP with HTTP Digest

SIP digest authentication is based on HTTP digest authentication used by HTTP communication with Web servers. It authenticates a client using a username (identity) and a password or secret key in a simple challenge-response authentication, applying cryptographic hash functions. Thus, the password is never sent in the clear. Nevertheless, there are some deficiencies in the usage of the HTTP Digest scheme, as it does not provide complete message integrity and cannot be applied to all messages. Here Transport Layer Security (TLS) kicks in, which can be used for signaling protection.

## B. SIP with TLS

The TLS protocol is the successor of the well-known Secure Socket Layer (SSL) protocol. It protects all communication on the transport layer for TCP connections against loss of integrity, confidentiality and against replay attacks. RFC3261 mandates the support of TLS for SIP proxies, redirect servers, and registrars to protect SIP signaling. Using TLS for User Agents (UAs), the SIP clients, is recommended. It provides integrated key-management with mutual authentication and secure key distribution. TLS is applicable hop-by-hop between UAs and proxies or between proxies. The SIP-Secure (SIPS) scheme defined in RFC3261 requires the usage of TLS to protect the signaling until the last proxy in the call flow. According to RFC3261 the last hop (from the proxy to the client) remains unprotected as a separate signaling connection is used for sending and receiving. This deficiency is being handled as part of RFC5626, describing an option to use client initiated connections also for the return signaling. RFC5923 is discussing a similar solution for the connection of two communicating proxies (cf. [6]).

## C. S/MIME to protect SIP message body data

The S/MIME standard is commonly used for encrypting and signing emails. RFC3261 recommends S/MIME to be used for end-to-end protection of SIP signaling message payloads following a similar approach as email. S/MIME within SIP supports authentication, integrity protection and confidentiality of signaling data. However, S/MIME is not widely used in current SIP deployments. One reason may be the overhead produced by S/MIME in terms of message size and complexity of parsing S/MIME protected content, which may be not acceptable in synchronous communication.

As new scenarios arise, the defined security measures within SIP do not always provide a solution. SIP is flexible with regard to extending the protocol. Therefore several security enhancements have already been standardized. Two approaches supporting identity management services needed in the different security levels are described in the following as prominent examples for extensions.

## D. Authenticated Identity Management

RFC4474 defines enhancements for SIP providing assertions for the user identity (Address of Record) valid in the domain the authentication server is responsible for to securely identify originators of SIP messages [3]. New header fields include a signature used for validating the identity and a reference to the certificate of the signer. As RFC4474 only addresses the request direction, a further standard has been defined: RFC4916 [4] describes the application of an authenticated identity service for the response direction.

Unfortunately, both documents do not provide a solution suitable for all relevant scenarios. They preferably work in adjacent domains. If, however, multiple administrative domains are to be traversed, the likelihood increases that a Session Border Controller SBC, i.e. a SIP proxy running on the border of a network, or a SIP proxy working as Back-to-Back user agent (B2B UA) is located on the signaling path. A B2B UA is a proxy that terminates the signaling traffic in both directions. This intermediate node may alter the SIP header or even the SIP body, hence destroying the signature of the authentication proxy. As the problem of a real end-to-end identity is not solved with the available solutions, further approaches are expected to be proposed.

## E. SIP/SAML

The Security Assertion Markup Language SAML is an XML extension for exchanging security information that has been developed by OASIS. SAML is a XML-based framework for creating and exchanging security information. When SIP requests are received by a server, there may be authorization requirements that are orthogonal to ascertaining the identity of the User Agent Client (UAC). Supplemental authorization information might allow the UAC to implement non-identity-based policies that depend on further attributes of the principal that originated a SIP request.

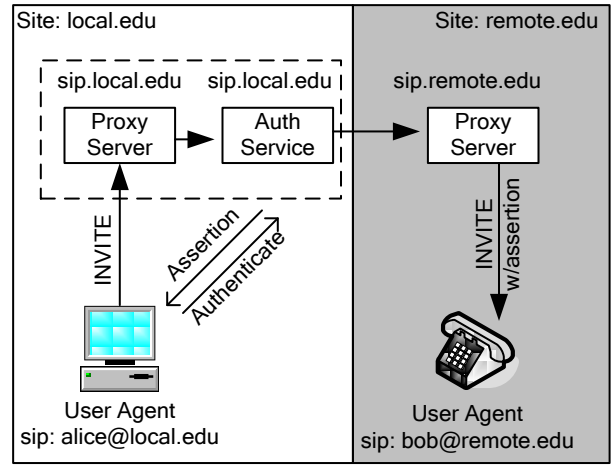

Figure 3. SAML Usage in SIP

An Internet draft [5] proposes a method for using SAML in collaboration with SIP to accommodate richer authorization mechanisms and enable trait-based authorization based on roles or traits instead of identity (see Figure 3). Moreover, it also defines how SAML assertions can be carried within SIP, which can be used end-to-end to vouch for a certain identity.

As stated at the beginning of this section, up to now SIP does not provide a universal identity management solution, which can be applied in every scenario. Therefore, the discussion within the IETF proceeds defining dedicated elements in the SIP messages, which may be altered by intermediate components, without scarifying the complete message identity. There already exist drafts describing potential solutions.

## V. USAGE AREA VOICE OVER IP COMMUNICATION: PROVIDING SIP IDENTITIES THROUGH ePA

The ePA user authentication can be applied in SIP-based VoIP telephony, either directly integrated into or adjacent to the SIP protocol

Different use cases for ePA-based user authentication pose different requirements on the technical solution as described in the following. The authentication can be performed towards the SIP service provider or toward the SIP communication peer (calleé), see Figure 4. Therefore, also different technical options for integrating ePA-based user authentication within SIP are needed.

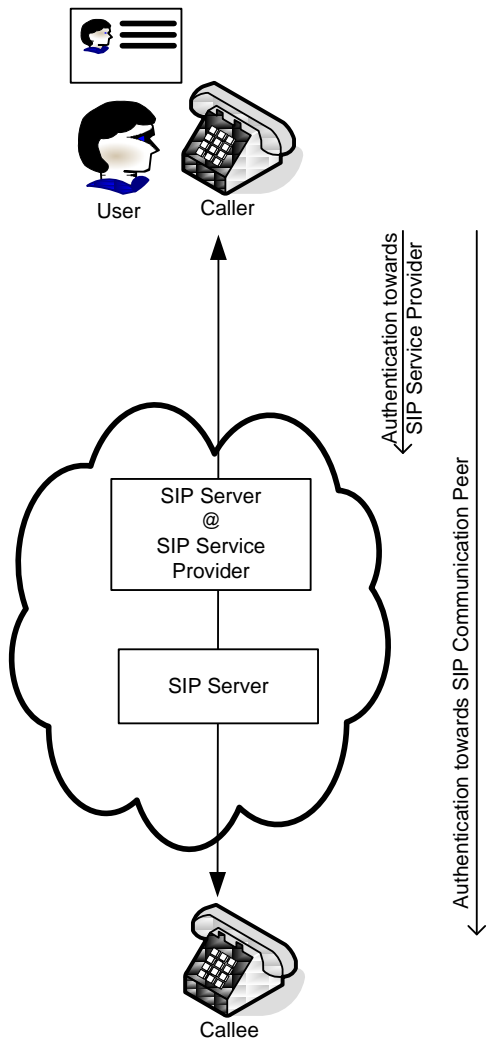

Figure 4. ePA-based VoIP/SIP User Authentication Scenarios

### A. Use Cases

User authentication within SIP protocol supported by the ePA functionality can be used to realize different application scenarios:

- Authentication of the user towards the SIP service provider for authorizing the use of SIP-based communication services. An ePA-based user authentication is performed when using SIP-based communication services.
- Authentication of the user towards the SIP service provider for bootstrapping security credentials needed for using SIP-based communication services. An ePA-based user authentication is performed as part of the establishment of a valid SIP-device configuration.
- Authentication of a user as part of registering with a communication service provider using a self-subscription Web portal. Here, the ePA user authentication is not used for technical authentication within the SIP application, but to establish the business relationship (contract) with a service provider. This may happen independently of SIP communication services.
- Authentication of the user towards an Identity Provider independently of the SIP signaling. The Identity provider issues an assertion (e.g., SAML assertion) that can be used with various communication services. Assertions may even be part of SIP.
- Authentication of the user towards the communication partner. Assured user identification or attributes of the user (as e.g., the user's age or even a pseudonym) can be provided to the communication peer. This information can be used by the communication peer for providing personalized communication services. For example, when calling a service hotline (merchant, public institution as a finance office), the calling user can be automatically identified resp. the user's attributes can be verified. When e.g., premium communication services are used, attributes as, e.g., the user's age can be verified before providing services.
- Media encryption: Session keys derived from ePA authentication may be used to encrypt the media stream towards the SIP service provider or the SIP communication peer. However, protection of VoIP communication besides authentication is out of scope of this paper.

Using ePA-based authentication towards the communication peer cannot be used in all of the described use cases. In particular in public SIP service offerings, SIP signaling is often terminated by the service provider's SIP server, so that specific SIP signaling may not reach the communication peer. This specifically applies to SIP headers, which may be added or removed by intermediaries.

The stated use cases vary concerning the following main requirements:

- SIP Protocol Integration: The ePA user authentication can be integrated within the SIP protocol itself, or it can be performed outside the SIP protocol for establishing a security context that may be used within SIP later on.
- Authentication Peer: The ePA authentication information can be performed towards a generic identity provider, a SIP service provider, or the communication peer.

- Frequency: An ePA-based user authentication can be performed only once to bootstrap SIP security configuration, or with each SIP session.

Therefore, different technical solutions for integration of ePA-based user authentication within VoIP are outlined in the next subsection.

*B. Integration of ePA-based Authentication within SIP*

Different possibilities exist for integrating ePA user authentication into SIP applications:

- Manual integration using an online Web authentication service: if a voice services requires identification, the caller connects to an associated website and uses the ePA Web Authentication mechanism. If authentication succeeds, a one time password is given to the caller that can be used for authentication to voice services, e.g., by entering the password using DTMF tones. This obviously requires a connection of the voice service provider with the ePA authentication infrastructure.

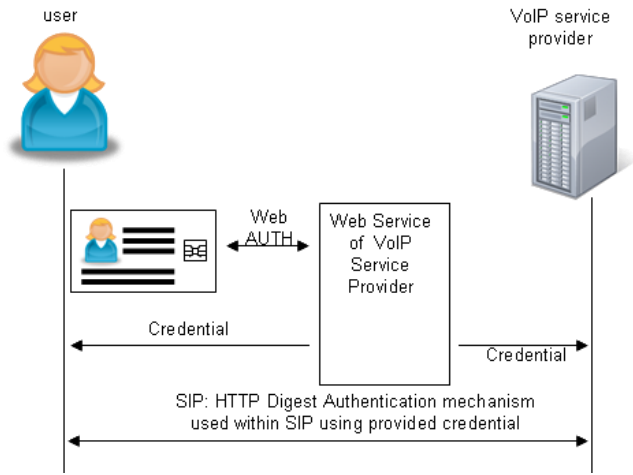

Figure 6.    ePA integration using HTTP digest

- Perform ePA client authentication within TLS security protocol protecting the SIP signaling channel towards the service provider, assumed the service provider possess the specific ePA server certificate. This would be transparent to the SIP protocol itself and completely relies on the usage of SIPS (SIP over TLS). It allows for authentication towards the SIP service provider.
- Authentication towards a separate identity provider issuing an identity assertion: A separate identity provider asserting identities is a flexible solution that allows fine-grain user control of revealed data. However, this approach requires both, the calleé and caller, to trust the identity provider. Involving a third party also poses privacy risks as the identity provider can collect caller data. This approach may be performed inband the SIP signaling as described for the SAML application in SIP.

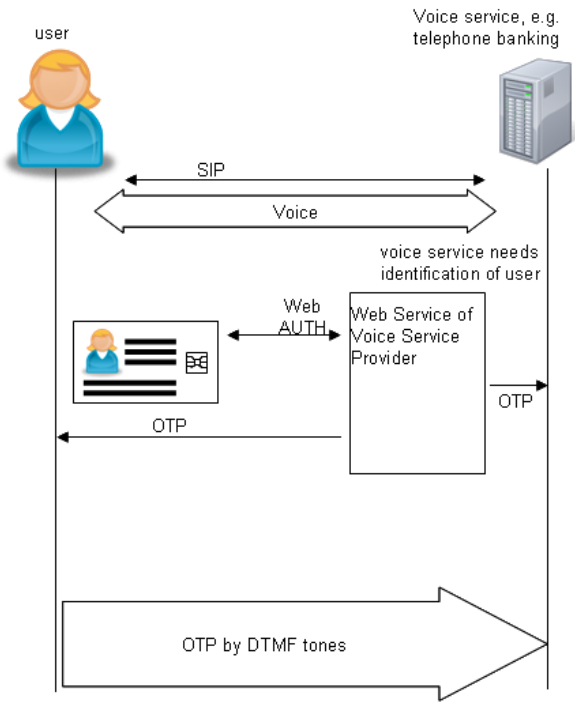

Figure 5.    Manual ePA Integration

- HTTP Digest within SIP: The ePA authentication requires a specific server certificate to activate the ePA client authentication. Hence an ePA Web authentication is performed first to obtain credentials that are used as HTTP Digest parameters within SIP. The same requirement as above applies to this example.

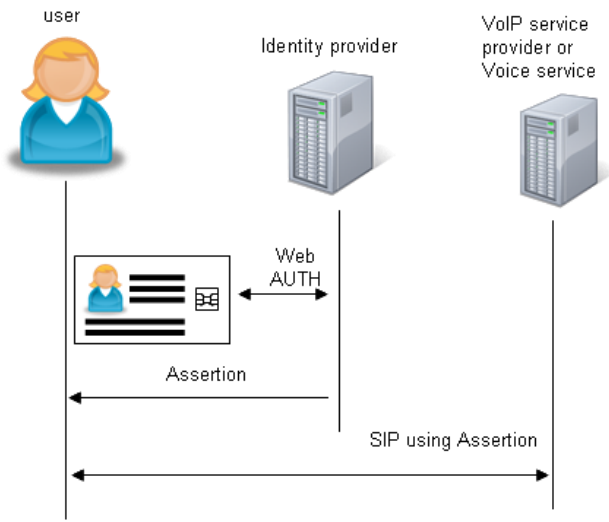

Figure 7. ePA integration using identity provider

- SIP with S/MIME: For end-to-end authentication, the integration of the ePA in SIP may use an additional part of the SIP message body. This part can be protected with S/MIME hence allows for integrity protection of the caller information through signing the S/MIME part with the callers private key. Moreover, if pseudonyms are used in the outer SIP header, the S/MIME part of the message may be used to transport the real identity of the caller in a privacy protected manner. This can be achieved by encrypting the S/MIME part using the calleé's public key (certificate). Note that this requires the availability of the calleé's certificate before establishing the call.

C. *Reality Check*

From a real-world perspective, the following use cases seem to be appropriate:

- ePA Authentication towards a Web Portal: The ePA-based Web authentication mechanisms is already available. Extensions to the SIP protocol are not necessary, as HTTP Digest authentication can be used as defined. Necessary is the integration of a preceding step binding the ePA Web Authentication to the HTTP Digest Authentication.
- ePA Authentication using an Identity Provider (SIP/SAML): The ePA-based authentication mechanisms is readily available. Extensions to the SIP protocol are already defined but it has to be considered, that they are still in draft status.
- ePA Authentication towards SIP Service provider: The easiest approach to integrate ePA authentication within SIP is to map the Web-based authentication on SIP, i.e. to use TLS for server authentication and to transport ePA authentication messages within SIP in the same way as over HTTP. Here, the SIP service provider needs an ePA terminal certificate to verify the user's identity. The SIP service provider may include verified user attributes in the SIP signaling towards other SIP service providers in case of multiple provider spanning connections, e.g., using the P-asserted identity extension (cf. [7]).
- ePA Authentication towards SIP Communication Peer: A direct integration within SIP signaling is difficult, as a SIP communication partner will usually not possess an ePA terminal certificate. The ePA user authentication follows a strict client/server principle, not a symmetric peer-to-peer like model as VoIP telephony. Also SIP signaling will often be terminated by the SIP service provider. Therefore a realistic approach seems to run an ePA-based user authentication towards a service/identity provider. The caller and the calleé may authenticate independently. Then known SIP security mechanisms can be used for end-to-end security (asserted identity, media encryption). The security association is, however, not established directly between the communicating entities, but by infrastructure nodes that have to be trusted.

VI. USAGE AREA HOME ENERGY MANAGEMENT

Allowing consumers to monitor their current energy usage level, even from remote, is one of the visions of the smart grid. The expectation is that energy usage monitoring results in a wiser use of energy in everyday life. Figure 8 shows a typical smart grid architecture.

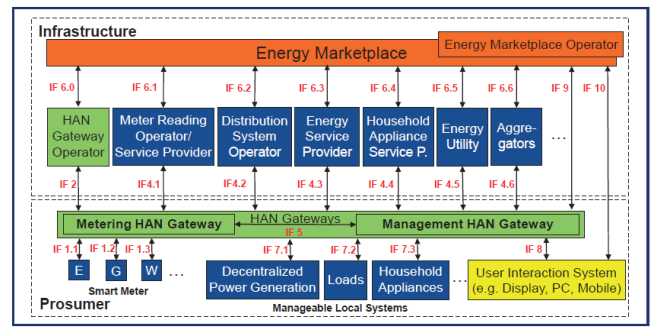

Figure 8. Smart Grid Reference Architecture from [9]

The architecture includes an "intelligent home" at the prosumer side (prosumer=energy PROducer + energy conSUMER), especially a HAN Gateway (Home Area Network Gateway) that exposes the available information on the energy consumption from the home to the outside, e.g., the Internet. To protect the smart home against attacks from the Internet, it is required to authenticate the communication (establishment) at the HAN Gateway. In the smart grid system, most communication partners of the HAN Gateway are already known and rather static. Secure identities are likely to already exist, for instance for certain control operations of or pricing information from the smart grid. However, secure identities for users accessing their HAN Gateway to monitor energy usage via Internet do not exist. Electronic identity cards can be used here to provide the secure identities needed.

*A. Use Cases*

The user can access information about its smart home (e.g. energy usage in the home) using a web browser. Several ways exist to implement this energy consumption web page:
1. The HAN Gateway runs a web server and provides the web page. The user authenticates towards the HAN Gateway using the ePA. However, this would require the HAN Gateway to implement an eID server. This means that the HAN gateway must comply to the very strict requirements defined in [10], which for example requires a hardware security module.
2. The utility provides the web page and polls the HAN Gateway for information. In this case, the user authenticates towards the utility using the ePA. The utility uses existing security associations that are independent of the ePA authentication to secure communication with the HAN Gateway. The prosumer could also use the ePA authentication towards the utility to establish or change contractual agreements as e.g. the selected tariff.

*B. Options for Integration ePA-based User Authentication*

Different possibilities exist for integrating ePA user authentication with HTTP access to the web page over a public network (Internet) providing information about the HAN Gateway:
- Direct use of web authentication (see section II)
- HTTP Digest: The HTTP protocol is used to establish a session between the HAN Gateway and the user. The ePA authentication requires a specific server certificate to activate the ePA client authentication. Hence an ePA Web authentication is performed first to obtain credentials that are used as HTTP Digest parameters within HTTP access towards the Web server (see Figure 9. ).

- Perform ePA client authentication within the TLS security protocol protecting the communication towards the web server, assuming the HAN Gateway possesses the specific ePA server certificate.
- Authentication towards a separate identity provider issuing an identity assertion (see Figure 10. ): A separate identity provider asserting identities is a flexible solution that allows fine-grained user control of revealed data. However, this approach requires both communication parties to trust the identity provider. In the home energy monitoring use case, the utility may for example be such a trusted party. Involving a third party also poses privacy risks as the identity provider can collect service usage information. Here, it may be possible to use SAML or Kerberos to issue a security token asserting the successful authentication towards the identity provider

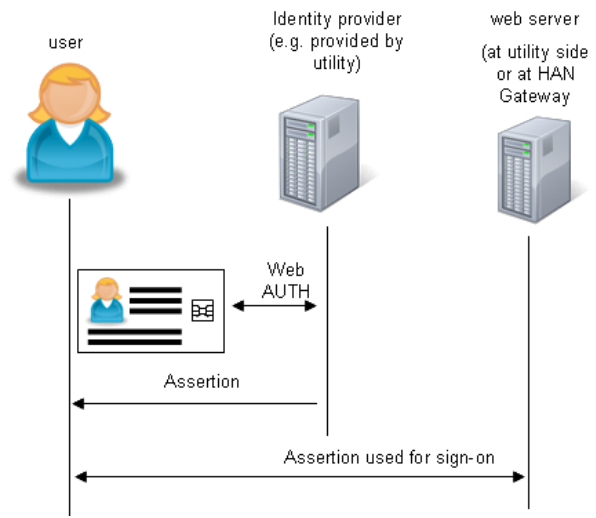

Figure 10. ePA integration using identity provider

*C. Reality Check*

From a real-world perspective, the following use cases could be realized:

- Direct use of web authentication: This integration approach seems to be only realistic if the utility provides the web site that the user accesses. In particular, a special server certificate is needed for the web server to interact with the ePA. Expecting each HAN Gateway to obtain such a server certificate seems unrealistic. Moreover, often HAN Gateways in a residential area will be connected via a Digital Subscriber Line (DSL) or power line communication and may therefore not be accessible by the prosumer directly (they do not have a public, static IP address, requiring services like MobileIP or DynDNS, to make them accessible from the public Internet).
- HTTP Digest: As with the approach above, this approach seems to be only realistic if the web site that the user accesses, is provided by the utility. However, this kind of session establishment involves an unnecessary indirection. This indirection may be justified if the HAN Gateway does not possess a fixed IP address, e.g., through connectivity via DSL. Thus address resolution for the HAN gateway is necessary. This may be achieved by having the HAN gateway permanently registered with a utility server, detecting IP address changes via the registration messages. This registration should be done using a secured connection between the HAN Gateway and the utility server. This utility server may then also perform the HTTP digest authentication mechanism and act as a proxy for remote access to the HAN Gateway.

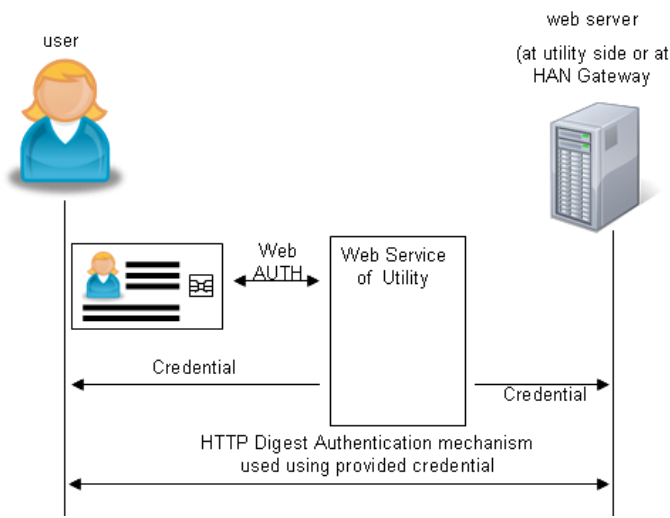

Figure 9. ePA integration using intermediate credential provider

- ePA Authentication towards a Web Portal or an Identity Provider (SAML): The utility can provide identities as it has a customer relationship with the user.

As the usage are home energy management is not yet mature, there is still the opportunity to embed ePA authentication into emerging protocols and standards.

## VII. USAGE AREA CHARGING SPOT ACCESS

Electric vehicles are becoming more and more important in the national and international strategies, e.g., to reduce $CO_2$ emissions. The integration of e-cars into the Smart Grid is still an evolving issue. Architectures like seen in Figure 11. are currently emerging.

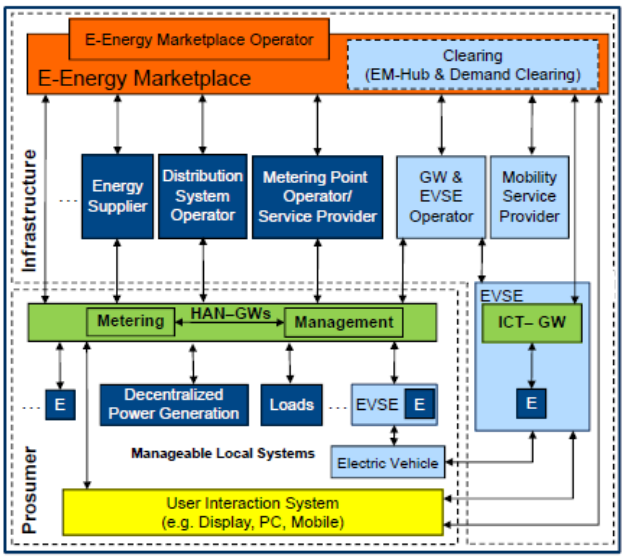

Figure 11. eMobility Extension of Smart Grid Reference Architecture from [9]

Authentication using the ePA can be useful for usage authorization at public charging spots. An authentication using a secure identity is important as the identity is used for accounting and billing. A user lock-in by different solutions of different vendors can be avoided if the already available ePA infrastructure is used for authentication. A charging provider could re-use an existing authentication infrastructure without having to issue separate authentication tokens. This has to be evaluated regarding potential privacy issues when re-using an authentication service revealing the user identity.

### A. Use Cases

The user authenticates towards a charging spot before the user's electric car is charged.

### B. Options for Integration ePA-based User Authentication

Different possibilities exist for integrating ePA user authentication as the user may authenticate directly at the charging spot or from inside the car. This requires the appropriate ePA interfaces on either the car or the charging spot. In the following, the focus is ePA authentication at the charging spot directly:

- Direct terminal authentication (see Section II): Each charging spot is acting as terminal for ePA authentication.
- HTTP Digest (see Figure 12. ): The charging spot uses HTTP to establish a session with the charging provider in the backend system. The ePA authentication requires a specific server certificate to activate the ePA client authentication. Hence an ePA Web authentication is performed first to obtain credentials that are used as HTTP Digest parameters.

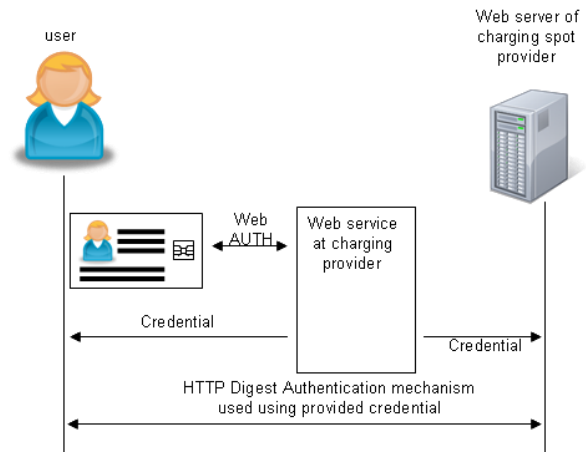

Figure 12. ePA integration using HTTP digest

- Perform ePA client authentication within the TLS security protocol protecting communication towards the charging service provider, where the charging provider possesses the specific ePA server certificate.
- Authentication towards a separate identity provider issuing an identity assertion (see Figure 13. ): A separate identity provider asserting identities is a flexible solution that allows fine-grained user control of revealed data. This approach allows easy integration with existing authentication solutions. However, this approach requires both the charging spot provider and the user to trust the identity provider. Involving a third party also poses privacy risks as the identity provider can collect mobility data. This approach may use for example SAML Assertions.or Kerberos tickets.

The identity used for authentication is the name of the user, potentially together with the user's address information. However, the current design of charging points often requires a contract ID instead of a real user name. The contract ID is linked with a name (resp. an associated database entry) in the backend system.

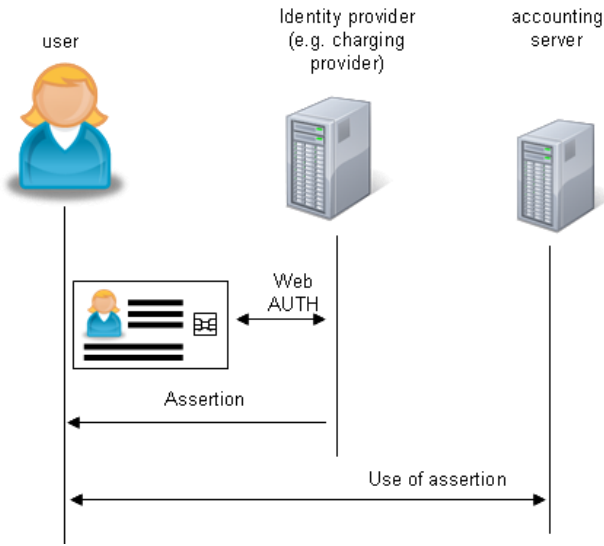

Figure 13.  ePA integration using identity provider

## C. Reality Check

From a real-world perspective, the following use cases could be realized, assuming that the charging spot provides the required interfaces for ePA interaction:

- Direct terminal authentication requires ePA server certificate for acting as a terminal as well as additional hardware [14], which may result in higher costs for charging spots.
- HTTP Digest: This approach could be implemented using available technology.
- ePA Authentication towards an Identity Provider (SAML): The ePA-based authentication mechanisms is available and the charging provider may serve as identity provider.

## VIII. OUTLOOK

Based on the use cases and potential solution options stated for the usage areas voice over IP communication, home energy management, and vehicle-to-grid integration the evolvement of identity card (e.g., ePA) based authentication in well established applications as well as in upcoming applications is highly expected. This expectation is supported by the fact that there is an available central authentication infrastructure, which can easily be used within a variety of services requiring authentication or some form of assertions of characteristics like the age of a person.

Moreover, from a technical point of view, it is also possible to load further applications onto the identity card, which enables further authentication schemes to be executed directly on the identity card supporting even more scenarios and usage areas by using the identity a universal secure transmission device of authentication credentials and applications. This would even enable further usage of the identity card in the curse of key management protocols, e.g., to agree on a session secret used to provide integrity or confidentiality or to simply serve as hardware based key generator for the key management protocol. Note, that the legal ramifications having multiple applications residing on one identity card as base for this option need to be further elaborated.

Regarding the direct utilization within SIP, further investigation into the enhancement of currently defined assertion or claim based service support to better utilize ePA functionality is recommended. Thanks to the SIP extensibility, new authentication services may be added without changing the base protocol. Some of the stated use cases in Section V apply the ePA for the initial security bootstrapping of SIP user environments. This option should be investigated more deeply, e.g., in the context of already established frameworks like the GBA (Generic Bootstrapping Architecture) or SACRED, (Securely Available Credentials, RFC 3760), as it provides a more generic bootstrapping option.

Regarding home energy management and vehicle-2-grid, embedding the use of identity cards into upcoming protocols and standards is an important issue. Moreover, as already stated the possibility to load own applications onto an identity card may provide even further solution spaces.

## IX. CONCLUSION

This paper gives on overview on different aspects of authentication and identity management for voice communication, home energy management as well as eMobility. For authentication and identity management, identity cards like the German "Elektronischer Personalausweis" (ePA) can be used. In Germany, a security infrastructure for identity management using the ePA is provided that is highly trusted and can be beneficial for protecting mature applications like voice communication as well as upcoming applications like home energy management and vehicle-2-grid. This paper presented a number of options for integrating an ePA-based user authentication in protocols typically used in voice over IP communication (SIP - Session Initiation Protocol), home energy management, and vehicle-2-grid. Identification and authentication are important building blocks in the protection of communication in all three usage areas. They may serve as a base for the exchange of further (session based) keying material which may be used for further protection of the session, e.g., for protection of integrity and confidentiality.